\documentstyle[11pt,epsf]{article} 
\input axodraw.sty
\def\lsim{\:\raisebox{-0.5ex}{$\stackrel{\textstyle<}{\sim}$}\:}

\newcommand{\newc}{\newcommand}
\newc{\pbi}{pb$^{-1}$}
\newc{\ie}{{\it i.e.} }
\newc{\ti}{\tilde}
\newc{\ra}{\rightarrow}
\newc{\ee}{$e^+e^-$\ }
\newc{\qq}{$q\bar{q}$\ }
\newc{\mm}{$\mu^+\mu^-$\ }
\newc{\taus}{$\tau^+\tau^-$\ }
\newc{\uu}{$u\bar{u}$\ }
\newc{\eeee}{$e^+e^-\ra e^+e^-$\ }
\newc{\eemm}{$e^+e^-\ra \mu^+\mu^-$\ }
\newc{\eett}{$e^+e^-\ra \tau^+\tau^-$\ }
\def\Rs{R \hspace{-0.38em}/\;}
\newc{\beq}{\begin{eqnarray}}
\newc{\eeq}{\end{eqnarray}}
\newc{\dqu}{\delta_{qu}}
\newc{\dqd}{\delta_{qd}}
\newc{\non}{\nonumber}
\newc{\noi}{\noindent}

\def\ib#1,#2,#3{       {\it ibid.\/ }{\bf #1} (19#2) #3}
\def\ap#1,#2,#3{       {\it Ann.~Phys.~(NY)\/ }{\bf #1} (19#2) #3}
\def\ijmp#1,#2,#3{     {\it Int.\ J.~Mod.\ Phys.\/ } {\bf A#1} (19#2) #3}
\def\mpl#1,#2,#3 {     {\it Mod.~Phys.~Lett.\/ } {\bf A#1} (19#2) #3}
\def\npb#1,#2,#3{       {\it Nucl.\ Phys.\/ }{\bf B#1} (19#2) #3}
\def\npps#1,#2,#3{     {\it Nucl.\ Phys.~B (Proc.~Suppl.)\/ }{\bf B#1}
                             (19#2) #3}
\def\plb#1,#2,#3{      {\it Phys.\ Lett.\/ }{\bf B#1} (19#2) #3}
\def\pr#1,#2,#3{       {\it Phys.\ Rev.\/ }{\bf #1} (19#2) #3}
\def\prd#1,#2,#3{      {\it Phys.\ Rev.\/ }{\bf D#1} (19#2) #3}
\def\prep#1,#2,#3{     {\it Phys.\ Rep.\/ }{\bf #1} (19#2) #3}
\def\prl#1,#2,#3{      {\it Phys.\ Rev.\ Lett.\/ }{\bf #1} (19#2) #3}
\def\pro#1,#2,#3{      {\it Prog.~Theor.\ Phys.\/ }{\bf #1} (19#2) #3}
\def\rmp#1,#2,#3{      {\it Rev.~Mod.~Phys.\/ }{\bf #1} (19#2) #3}
\def\sp#1,#2,#3{       {\it Sov.~Phys.~Usp.\/ }{\bf #1} (19#2) #3}
\def\zpc#1,#2,#3{      {\it Z.~Phys.\/ }{\bf C#1} (19#2) #3}
\def\appb#1,#2,#3{     {\it Acta Phys.\ Polon.\/ }{\bf B#1} (19#2) #3}

\begin{document}
\hfill IFT-22-97
\begin{center}
{\large \bf{SLEPTONS IN $R$-PARITY VIOLATING SUSY\footnote{Presented at 
the XXI School of Theoretical Physics ``Recent Progress in Theory and 
Phenomenology of Fundamental Interactions'', Ustro\'n, Poland, September 1997.
 }}}

\vspace{8mm}
{\large {\bf Jan Kalinowski}\\
Institute of Theoretical Physics, Ho\.za 69, PL-00681 Warsaw}
\end{center}

\vspace{1cm}
\begin{abstract}
In $R$-parity violating SUSY models  sleptons can be produced
singly in \ee and \qq collisions. The formation of slepton
resonances at LEP2 or Tevatron at current energies is an
exciting possibility. Existing LEP2 and
Tevatron data can be exploited to look for sleptons and , if
unsuccessful, to derive bounds on the Yukawa
couplings of sleptons to quark and lepton pairs.
\end{abstract}
\section{Introduction}
This year has witnessed an increase of interest in the
$R$-parity violating supersymmetric model. It has been triggered
by observation at HERA of a number of events at high $Q^2$, high
$x$ in $e^+p$ scattering \cite{data}. Although the experimental
situation remains unsettled, the supersymmetry with broken
$R$-parity has been put forward as a possible explanation of these
events \cite{spi}.

In the usual formulation, the minimal supersymmetric extension (MSSM)
of the Standard Model (SM) is defined by the superpotential 
\begin{equation}
  W_R=Y_{ij}^e L_iH_1 E^c_j + Y_{ij}^d Q_i H_1D^c_j
  +Y_{ij}^uQ_iH_2U^c_j +  \mu H_1 H_2 \label{Rcons}
\end{equation}
which respects a  
discrete multiplicative symmetry, $R$-parity, 
which can be defined as \cite{FF}
\begin{equation}
             R_p=(-1)^{3B+L+2S}
\end{equation} 
where $B$, $L$ and $S$ denote the baryon and lepton number, and
the spin of the particle: all Higgs particles and SM 
fermions and bosons have $R_p=+1$, and their superpartners
have $R_p=-1$.

  In eq.\ (\ref{Rcons})  the standard notation is used for the
left-handed doublets of leptons ($L$) and quarks ($Q$), the
right-handed singlets of charged leptons ($E$) and down-type
quarks ($D$), and for the Higgs doublets which couple to the
down ($H_1$) and up quarks ($H_2$);  the indices $i,j$ denote
the generations, and a summation is understood, $Y^f_{ij}$ are
Yukawa couplings and $\mu$ is the Higgs mixing mass parameter.

Because of $R_p$ conservation, the interaction lagrangian
derived from $W_R$ contains terms in which 
the supersymmetric partners appear only in pairs. Therefore 
superpartners can be produced only in pairs in collisions and decays
of particles, and the lightest supersymmetric particle (LSP) is stable. 

However, there is no theoretical motivation for imposing $R_p$
since 
gauge and Lorentz symmetries allow for additional
terms in the superpotential 
\begin{equation}
  W_{\Rs}=\lambda_{ijk}L_iL_jE^c_k + \lambda'_{ijk}L_iQ_jD^c_k
  +\lambda''_{ijk} U^c_iD^c_jD^c_k\label{Rviol} +
   \epsilon_i L_i H_2 
\end{equation} 
which break explicitely the $R$-parity \cite{WSY}. If the 
Yukawa couplings $\lambda$, $\lambda'$, $\lambda''$
and/or dimensionful mass parameters $\epsilon$ are
present, the model has distinct features: superpartners can be
produced singly and the LSP is not stable.  Because of
anti-commutativity of the superfields, $\lambda_{ijk}$ can be chosen to
be non-vanishing only for $i < j$ and $\lambda''_{ijk}$ for $j<k$.
Therefore for three generations of fermions, $W_{\Rs}$ contains
additional 48 new parameters beyond those in eq.~(\ref{Rcons}). Note
that at least two different generations are coupled in the purely
leptonic or purely hadronic operators.

The couplings $\lambda$, $\lambda'$ and $\epsilon$ violate lepton
number ($L$), whereas $\lambda''$ couplings violate baryon number
($B$), and thus can possibly lead to fast proton decay if both
types of couplings are present.  In the usual
formulation of the MSSM they are forbidden by $R$-parity and
the proton is stable.  
However other discrete symmetries can allow for a stable proton and
$R$-parity violation at the same time. For example,
baryon-parity (defined as $-1$ for quarks, and $+1$ for leptons and
Higgs bosons) implies $\lambda''=0$. In this case only lepton number
is broken, which suffices to ensure proton stability. Lepton-number 
violating operators can also provide new ways to generate neutrino masses. 

From the
grand unification and string theory points of view, both types of
models, $R_p$ conserving or violating, have been constructed with no
preference for either of the two \cite{D1}.  Since they lead to very
different phenomenology, both models should be searched for
experimentally.  The MSSM with $R_p$-conservation
has been extensively studied phenomenologically and experimentally.
Here we will consider the MSSM with broken $R_p$ with the most
general trilinear 
terms in eq.~(\ref{Rviol}) that violate $L$ but conserve $B$.

In the Lagrangian the  $\lambda$ and $\lambda'$ parts of
the Yukawa interactions have the following form: 
\begin{eqnarray} {\cal
  L}_{\Rs}&=&\lambda_{ijk}\left[ \ti{\nu}^j_L\bar{e}^k_Re^i_L
+\overline{\ti{e}}^k_R (\bar{e}^i_L)^c\nu^j_L
+\ti{e}^i_L\bar{e}^k_R\nu^j_L\right. \non \\
&&\left. \mbox{~~~~~~~~~}
-\ti{\nu}^i_L\bar{e}^k_Re^j_L-\overline{\ti{e}}^k_R (\bar{e}^j_L)^c\nu^i_L
-\ti{e}^j_L\bar{e}^k_R\nu^i_L \right] + h.c.  \nonumber \\ 
&+&\lambda'_{ijk}\left[( \ti{u}^j_L\bar{d}^k_Re^i_L
+\overline{\ti{d}}^k_R(\bar{e}^i_L)^cu^j_L +\ti{e}^i_L\bar{d}^k_Ru^j_L
)\right. \nonumber \\ &&\left.
\mbox{~~~~~~~~~}-(\ti{\nu}^i_L\bar{d}^k_Rd^j_L+\ti{d}^j_L\bar{d}^k_R\nu^i_L
+\overline{\ti{d}}^k_R(\bar{\nu}^i_L)^cd^j_L) \right] + h.c.\mbox{~~~}
\label{effl}
\end{eqnarray} 
where $u_i$ and $d_i$ stand for $u$- and $d$-type quarks, $e_i$ and
$\nu_i$ denote the charged leptons and neutrinos of the $i$-th
generation, respectively; the scalar partners are denoted by a tilde.
In the $\lambda'$ terms, the up (s)quarks in the first parentheses
and/or down (s)quarks in the second may be Cabibbo rotated in the
mass-eigenstate basis.  As we will discuss mainly sneutrino induced
processes, we will assume the basis in which only the up sector is
mixed, $i.e.$ the $NDD^c$ is diagonal. 

 The
possibility of some $\lambda$ and $\lambda'$ couplings to be
non-zero opens many interesting processes at
current and future colliders in which this scenario can be
explored. Since from low-energy
experiments the 
Yukawa couplings are expected to be small, indirect effects due to
$t/u$-channel exchanges of sfermions in collisions of leptons and
hadrons can be difficult to observe.  However, the direct formation of
sfermion resonances in the $s$-channel processes can produce
measurable effects.  For example, squarks could be produced as
$s$-channel resonances in lepton-hadron collisions at HERA.  In fact,
recent high $Q^2$, high $x$ events at HERA have been analyzed in this
context; higher statistics however is needed to draw definite
conclusions.  Sleptons on the other hand could be produced as
$s$-channel resonances in lepton-lepton and hadron-hadron collisions,
 and could decay to leptonic or hadronic final states in addition to
$R$-parity conserving modes.

Note that since in SUSY GUT scenarios sleptons are generally expected
to be lighter than squarks, sleptons may show up at LEP2 and/or
Tevatron even if squarks are beyond the kinematical reach of HERA.
Therefore we will consider the possible effects of $s$-channel
slepton resonance production in $e^+e^-$ collisions
\begin{eqnarray}
& & e^+e^- \rightarrow \tilde{\nu}\rightarrow \ell^+\ell^- \label{elel} \\ 
& & e^+e^- \rightarrow \tilde{\nu}\rightarrow q\bar{q} \label{ququ}
\end{eqnarray}
and in $p\bar{p}$ collisions 
\begin{eqnarray}
&& p\bar{p} \rightarrow \tilde{\nu}\rightarrow \ell^+\ell^-
\label{drel} \\
&& p\bar{p} \rightarrow \tilde{\ell}^+\rightarrow \ell^+\nu \label{lnu}  
\end{eqnarray}The results presented here
have been obtained in collaboration with H. Spiesberger, R. R\"uckl
and P. Zerwas \cite{snu,tev}.

\section{Sfermion Exchanges in $f\bar{f}'\rightarrow F\bar{F}'$ Processes}
Once produced, sleptons can decay via either the $R_p$ violating
Yukawa 
or the $R_p$ conserving gauge couplings. In the latter case the
decay proceeds in a cascade process which involves standard and 
supersymmetric particles in the intermediate states and with the
$R_p$ violating coupling 
appearing at
 the end of the cascade. Such decay processes lead in general
to  multibody final 
states and depend on many unknown SUSY parameters.  
In the former case, the final state is a two-body state
(with two visible particles, eqs.~(\ref{elel})-(\ref{drel}),  or one
visible particle and a missing 
momentum, eq.~(\ref{lnu})) which depends only on a limited
number of parameters and which is very easy to 
analyse experimentally. Therefore we will consider sleptons that
are produced and decay via $\lambda$ and/or $\lambda'$ couplings.

Let us consider first a generic
two-body process $f\bar{f}'\rightarrow F\bar{F}'$ which in the Standard
Model can proceed via $s$- and/or $t$-channel gauge boson exchange
($\gamma$, $Z$, or $W$; for light fermions the Higgs boson exchange is
negligible), as shown in Fig.~1. Turning on the $\lambda$ and
$\lambda'$ couplings, there are additional contributions due to 
sfermions which can contribute via $s$-, $t$-, and/or $u$-channel exchange
processes, Fig.~1. The differential cross section in the $f\bar{f}'$ 
rest frame 
can be written in terms of helicity amplitudes as follows
\begin{eqnarray}
&&\frac{\mbox{d}\sigma}{\mbox{d}\cos\theta} (f\bar{f}' \ra F\bar{F}') 
= A_c
\frac{\pi\alpha^2s}{8}
\Bigl\{ 4\left[|f^t_{LL}|^2+|f^t_{RR}|^2\right]\non \\
&& \mbox{~~~~~~~~~} +  
 (1-\cos\theta)^2\left[|f^s_{LL}|^2+|f^s_{RR}|^2\right]
\non\\
&&\mbox{~~~~~~~~~} + (1+\cos\theta)^2 
\left[ |f^s_{LR}|^2 + |f^s_{RL}|^2 +
  |f^t_{LR}|^2 + |f^t_{RL}|^2 
\right.\non \\ 
&&\left. \mbox{~~~~~~~~~~~~~~~~~~~~~~~~~~}
+ 2\mbox{Re}(f^{s\;*}_{LR}\,f^t_{LR} + 
     f^{s\;*}_{RL}\,f^t_{RL}) \right] 
   \Bigr\}
\label{dsigdcos}
\end{eqnarray}
where $A_c$ is the appropriate color factor. 
To simplify notations we
have defined the indices $L,R$ to denote the helicities of the {\it
incoming fermion $f$} (first index) and the {\it outgoing antifermion $F'$}
(second index).  
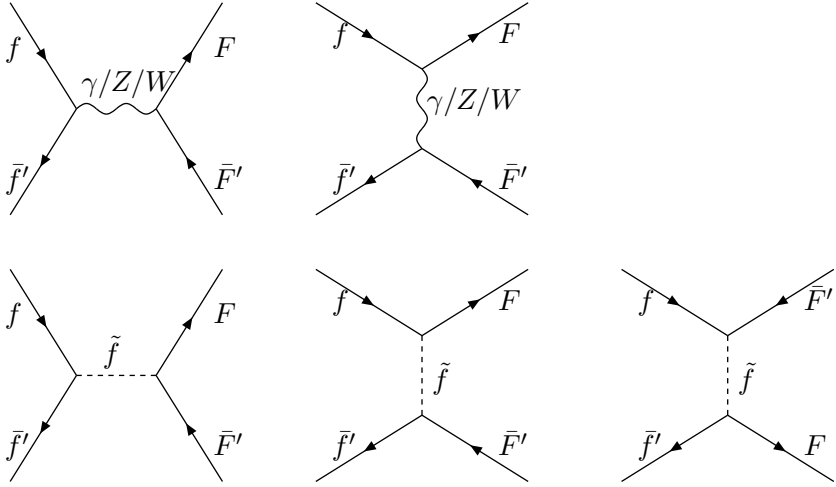
\begin{figure}[htbp]
%
\begin{picture}(100,100)(0,0)
\ArrowLine(10,90)(35,50)
\ArrowLine(35,50)(10,10)
\Photon(35,50)(65,50){2}{2}
\ArrowLine(90,10)(65,50)
\ArrowLine(65,50)(90,90)
\put(8,20){$\bar{f}'$}
\put(8,70){$f$}
\put(37,57){$\gamma/Z/W$}
\put(87,20){$\bar{F}'$}
\put(87,70){$F$}
\end{picture}
%
\begin{picture}(100,100)(-12,0)
\ArrowLine(10,90)(50,65)
\ArrowLine(50,65)(90,90)
\Photon(50,65)(50,35){2}{2}
\ArrowLine(90,10)(50,35)
\ArrowLine(50,35)(10,10)
\put(16,20){$\bar{f}'$}
\put(16,75){$f$}
\put(52,50){$\gamma/Z/W$}
\put(79,20){$\bar{F}'$}
\put(79,75){$F$}
\end{picture}
\\
%
\begin{picture}(100,100)(0,0)
\ArrowLine(10,90)(35,50)
\ArrowLine(35,50)(10,10)
\DashLine(35,50)(65,50){2}
\ArrowLine(90,10)(65,50)
\ArrowLine(65,50)(90,90)
\put(8,20){$\bar{f}'$}
\put(8,70){$f$}
\put(45,55){$\tilde{f}$}
\put(87,20){$\bar{F}'$}
\put(87,70){$F$}
\end{picture}
%
\begin{picture}(100,100)(-12,0)
\ArrowLine(10,90)(50,65)
\ArrowLine(50,65)(90,90)
\DashLine(50,65)(50,35){2}
\ArrowLine(90,10)(50,35)
\ArrowLine(50,35)(10,10)
\put(16,20){$\bar{f}'$}
\put(16,75){$f$}
\put(54,45){$\tilde{f}$}
\put(79,20){$\bar{F}'$}
\put(79,75){$F$}
\end{picture}
%
\begin{picture}(100,100)(-24,0)
\ArrowLine(10,90)(50,65)
\ArrowLine(90,90)(50,65)
\DashLine(50,65)(50,35){2}
\ArrowLine(50,35)(90,10)
\ArrowLine(50,35)(10,10)
\put(16,20){$\bar{f}'$}
\put(16,75){$f$}
\put(54,45){$\tilde{f}$}
\put(79,20){$F$}
\put(79,75){$\bar{F}'$}
\end{picture}
\caption{\label{baba}  Generic Feynman diagrams for 
  \protect$f\bar{f}'\rightarrow F\bar{F}'$
  scattering including $s$- and $t$-channel exchange of $\gamma/Z/W$, 
  and $s$-, $t$- and $u$-channel exchange of sfermion $\tilde{f}$.}
\end{figure}
The $s$- and $t$-channel $\gamma,Z,W$ amplitudes in the Standard
Model involve the coupling of vector currents. On the other hand
the sfermion exchange
is described by scalar couplings. However, by performing appropriate Fierz
transformations, 
\beq 
(\bar{f}_R f'_L)(\bar{F}_L F'_R) \ra
-\frac{1}{2}(\bar{f}_R\gamma_{\mu} F'_R)(\bar{F}_L\gamma_{\mu}f'_L)
\eeq 
for the field operators, 
the $s$-channel $\ti{f}$ exchange amplitudes  can 
be rewritten as $t$-channel vector amplitudes, and
$t/u$-channel $\ti{f}$ exchange amplitudes as $s$-channel vector
amplitudes. Therefore it is easy to see that  helicities of the
incoming antifermion and the 
outgoing fermion are fixed by the $\gamma_5$ invariance of the vector
interactions: they are opposite to the helicities of the fermionic 
partner in $s$-channel amplitudes and the same in $t$-channel
amplitudes.
 
In performing the Fierz transformation attention should be paid
to the relative signs of 
the SM and sfermion exchange amplitudes. We find that   
the $t$- and $u$-channel sfermion
contributions enter with the opposite signs due to different
ordering of fermion operators in 
the Wick reduction \cite{lq}. 

The independent $s$-channel amplitudes $f^s_{ij}$ ($i,j=L,R$)
can  be written as follows 
\newcommand{\half}{\frac{1}{2}}
\beq 
f^s_{ij} &=& \frac{Q^s_{ij}}{s} 
 +\half\; \frac{G^t_{ij}/e^2}{t-m^2_{\ti{f}}}
  -\half\; \frac{G^u_{ij}/e^2}{u-m^2_{\ti{f}}} \label{helamps}
\eeq
where $s=(p_f+p_{\bar{f}'})^2$, $\sqrt{s}$ is the center-of-mass
energy of the $f\bar{f}'$ system,
$t=(p_f-p_F)^2=-s(1-\cos\theta)/2$, and
$u=(p_f-p_{\bar{F}'})^2=-s(1+\cos\theta)/2$. 
Similarly, the $t$-channel exchange amplitudes $f^t_{ij}$ 
read
\begin{eqnarray}
f^t_{ij} &=& \frac{Q^t_{LR}}{t} 
+\half\frac{G^s_{ij}/e^2}{s-m^2_{\ti{f}}+i\Gamma_{\ti{f}} m_{\ti{f}}}
\label{helampt}
\end{eqnarray} 
The parameters $m_{\ti{f}}$ and $\Gamma_{\ti{f}}$ are the mass 
and width of the
exchanged sfermion $\ti{f}$ ($\tilde{f}$ is a generic notation
of the exchanged sfermion, not necessarily the superpartner of $f$).  
 The generalized SM charges $Q^{s,t}_{ij}$  for gauge
boson exchanges and the factors $G^{s,t,u}_{ij}$ in terms of
Yukawa couplings of the exchanged sfermions will be given when
specific reactions are discussed. In processes involving
$\gamma$ and $Z$ exchanges, the generalized charges $Q_{ij}$
depend on the momentum transfers and their signs  
determine the interference pattern of SM  with
sfermion exchange terms.

\section{Indirect Bounds on the Yukawa Couplings}
The masses and Yukawa couplings of sfermions are not predicted
by theory. 
At energies much lower than the sparticle masses, $R$-parity breaking
interactions can be formulated in terms of effective four-fermion contact
interactions.  These operators will in general mediate $L$ violating
processes and FCNC processes.  Since the existing data are
consistent with the SM,  stringent
constraints on the Yukawa couplings and masses can be derived \cite{limits}.  
However, if only some of the terms
with a particular generation structure are present in eq.\ 
(\ref{effl}), then the effective four-fermion Lagrangian is not strongly 
constrained.  Similarly, the couplings can be arranged such
that there are no other sources of FCNC interactions than CKM mixing
in the quark sector. Below we will consider the following two 
scenarios:
\\[1mm]
($i$) one single Yukawa coupling is large, all the other couplings
are small and thus neglected; \\[1mm]
($ii$) two Yukawa couplings which violate {\it one and the same}
lepton flavor are large, all the others are neglected.

Since theoretically the third-generation sfermions are expected
lighter than the first two and, due to large top quark
mass, the violation of the third-generation lepton-flavor might
be expected maximal, we will concentrate on possible effects
generated by $\tilde{\tau}$ and $\tilde{\nu}_{\tau}$, $i.e.$ we
are concerned with $\lambda_{i3i}$ and $\lambda'_{3jk}$
couplings.
In these cases low-energy experiments are not restrictive and
typically allow for couplings $\lambda \lsim 0.1\times$($\ti{m}$/200
GeV), where $\ti{m}$ is the mass scale of the sparticles participating
in the process.

\begin{figure}[htbp]
\begin{picture}(100,100)(0,0)
\ArrowLine(10,60)(40,60)
\ArrowLine(40,60)(70,90)
\Photon(40,60)(60,40){2}{2}
\ArrowLine(60,40)(90,70)
\ArrowLine(90,10)(60,40)
\put(5,62){ $\tau^-$ }
\put(93,67){ $e^-$}
\put(51,51){ $W$}
\put(73,87){$\nu_\tau$}
\put(93,7){$\nu_e$}
\end{picture}
\begin{picture}(100,100)(-12,0)
%
\ArrowLine(10,60)(40,60)
\ArrowLine(70,90)(40,60)
\DashLine(40,60)(60,40){2}
\ArrowLine(60,40)(90,70)
\ArrowLine(60,40)(90,10)
\put(5,62){ $\tau^-$ }
\put(93,67){ $e^-$}
\put(51,51){ $\tilde{e}$}
\put(73,87){$\nu_e$}
\put(93,7){$\nu_\tau$}
\end{picture}
\caption{Tau decay via the SM $W$ exchange, and via the
\protect$\ti{\nu}_{\tau}$ due to the  \protect$L_1L_3E^c_1$ operator.} 
\end{figure}
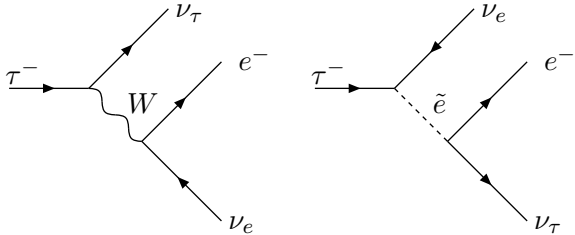
Let us consider a specific example. The operator
$\lambda_{131}L_1L_3E^c_1$ can contribute to the $\tau$
leptonic decay process $\tau\ra e\nu\bar{\nu}$ via the diagram in
Fig.~2. 
After Fierz transformation the sneutrino exchange diagram has the
same structure as the  SM $W$ exchange  
and thus leads to an apparent shift in the Fermi constant for tau decays.
The ratio $R_\tau\equiv\Gamma(\tau\ra e\nu{\bar\nu})/\Gamma
(\tau\ra\mu\nu{\bar\nu})$ relative to the
SM contribution is then modified \cite{barger}
\beq
R_\tau=R_\tau(SM)\left[1+2\frac{M_W^2}{g^2}
\left(\frac{|\lambda_{131}|^2}{{\tilde
m}^2({\tilde e}_R)}\right)\right].
\eeq
Using the experimental value \cite{pdb} we obtain \cite{snu} the bound
\beq
|\lambda_{131}| < 0.08\, \left(\frac{{\tilde
m}({\tilde e}_R)}{200~\mbox{GeV}}\right) 
\eeq
which is given in
Table 1.

Other limits relevant for $\lambda_{i3i}$
and $\lambda'_{3jk}$, derived by assuming only one non-vanishing
coupling at a time, are summarized in Table~\ref{tablam}.
\begin{table}[htbp]
\begin{center}
\footnotesize\rm
\caption{Low-energy limits for the couplings 
\protect$\lambda_{i3i}$
($i=1,2$) and \protect$\lambda'_{3jk}$ ($j=1,2$, $k=1,2,3$)  
assuming the relevant sfermion masses \protect$\ti{m}=200$ GeV. 
They are derived from 
(a) \protect$\Gamma(\tau\ra e\nu\bar{\nu})/ 
   \Gamma(\tau\ra\mu\nu\bar{\nu})$ \protect\cite{snu}; 
(b) \protect$\Gamma(\tau\ra e\nu\bar{\nu})/ 
   \Gamma(\mu\ra e\nu\bar{\nu})$ \protect\cite{barger}; 
(c) $K\ra \pi\nu\nu$ \protect\cite{agashe}; 
(d) $D\bar{D}$ mixing \protect\cite{D1}; 
(e) $\tau\ra \pi\nu$ \protect\cite{deb}.}  
\begin{tabular}{|c|cccc|}
\hline
 \rule{0mm}{5mm}
coupling&  $\lambda_{131}$& $\lambda_{232}$& $\lambda'_{3jk}$& 
$\lambda'_{31k}$  \\[1mm]
\hline
Low-energy  & 0.08$^a$& 0.08$^b$ & 0.024$^c$ & 0.32$^e$ \\ 
limit       &         &          & 0.34$^d$  & \\
\hline
\end{tabular}
\label{tablam}
\end{center} 
\end{table}
The limit (d) for $\lambda'_{3jk}<0.34$ is derived assuming the CKM
mixing due to absolute mixing in the up-quark sector only ($NDD^c$
diagonal); if the CKM mixing is due to absolute mixing in the
down-quark sector ($EQD^c$ diagonal), more stricter bound (c) of 0.024
applies.  In summary, present low-energy data allow
$\lambda_{i3i}\lsim 0.08$, and $\lambda_{i3i}\lambda'_{311}\lsim
(0.05)^2$, even in the limit (c).

\section{Sneutrinos in \ee Scattering }
In \ee scattering sneutrinos can be produced in the $s$-channel
and sleptons exchanged in the $t$ or $u$ channels, 
leading to a number of different signatures depending on the
assumed scenario. 
If only $\lambda_{131}\ne 0$, the tau sneutrino $\ti{f}=\ti{\nu}_{\tau}$ can
contribute to Bhabha scattering via $s$- and $t$-channel
exchanges, and the electron sneutrino $\ti{f}=\ti{\nu}_e$ in the
process   $e^+e^-\ra\tau^+\tau^-$ can be exchanged in the 
$t$-channel. Assuming in addition $\lambda_{232}\ne0$, also muon
pair production, $e^+e^-\ra\mu^+\mu^-$, can be mediated by the
$s$-channel $\ti{\nu}_{\tau}$ resonance.  
Taking $\lambda'_{3jk}\ne0$ would lead to $s$-channel
$\ti{\nu}_{\tau}$ contribution in hadronic processes $e^+e^-\ra
q_j\bar{q}_k$. 
We will consider these cases below. Note that apart from
$R$-parity violating decays, the $\ti{\nu}_{\tau}$ can also
decay via $R$-parity conserving modes; such decays have already been
discussed in the literature \cite{dl}. On the other hand,
$\ti{\tau}$ slepton in $e^+e^-$ collisions can only contribute
via $t/u$-channels to the neutrino-pair production cross section, which for
couplings considered here is below 1 \%.  \\[1mm]
\noindent (a) {\it Bhabha scattering:} 
The differential cross section for Bhabha scattering is given by
eq.~(\ref{dsigdcos}) with $A_c=1$, and the SM generalized charges in
helicity amplitudes are as follows
\begin{eqnarray}
&&Q^s_{ij} = 1+ g^e_i
g^e_{-j}\frac{s}{s-m^2_Z+i\Gamma_Zm_Z}\label{opposite} \\
&&Q^t_{ij} = 1+ g^e_i g^e_{-j}\frac{t}{t-m^2_Z} \non
\end{eqnarray}
The subscript $-j$ means that the helicity index is opposite to
$j$ because in eqs.~(\ref{helamps},\ref{helampt}) the outgoing
positron with the 
helicity $L(R)$ couples with the charge   $g^e_R (g^e_L)$, where the
  left/right
$Z$ charges  of the fermion $f$ are defined as
\begin{eqnarray}
g^f_L=(\frac{\sqrt{2}G_{\mu}m^2_Z}{\pi\alpha})^{1/2}
(I_3^f-s^2_W Q^f), \mbox{~~~~~~~} 
g^f_R=(\frac{\sqrt{2}G_{\mu}m^2_Z}{\pi\alpha})^{1/2}
(  {} -s^2_W Q^f)\non 
\end{eqnarray}
The $R_p$ violating sneutrino contributions are given in terms of the 
factors $G_{ij}$  as follows
\begin{eqnarray}
G^s_{LL}=G^s_{RR}=G^t_{LL}=G^t_{RR}=(\lambda_{131})^2
\end{eqnarray}
with all other $G_{ij}=0$. Note that the $s$-channel ($t$-) sneutrino
exchange interferes with the $t$-channel ($s$-) $\gamma,Z$
exchanges. 
\\[1mm]
\noindent (b) {\it Muon-pair production:} The
SM generalized charges $Q^s_{ij}$ are given by
eq.~(\ref{opposite}). 
Since the $t$-channel $\gamma,Z$ and $\ti{\nu}_{\tau}$ exchanges
are absent, $Q^t_{ij}=0$, $G^t_{ij}=0$, the $s$-channel
sneutrino exchange given by   
\begin{eqnarray}
G^s_{LL}=G^s_{RR}=\lambda_{131}\lambda_{232}, 
\mbox{~~~~all~other~~~}G_{ij}=0
\end{eqnarray}
does not interfere with the SM processes.
\\[1mm]
\noindent (c) {\it Tau-pair production:} 
This process can receive only
the $t$-channel exchange of $\ti{\nu}_e$ with 
\begin{eqnarray}
G^t_{RR}=(\lambda_{131})^2, \mbox{~~~~all~other~~~} G_{ij}=0 
\end{eqnarray}  
which will interfere with the SM $\gamma,Z$ $s$-channel
processes with $Q^s_{ij}$ given by eq.~(\ref{opposite}).\\[1mm] 
\noindent (d) {\it $e^+e^-$ annihilation to hadrons:}
The up-type quark-pair production is not affected by sneutrino
processes, as can be easily seen from the general structure of
couplings in eq.~(\ref{effl}). On the other hand, for the
down-type quark-pair production, $e^+e^-\ra d_k\bar{d}_k$, 
the differential cross section is given by
eq.~(\ref{dsigdcos}), however with the color factor $A_c=3$.  
In this case the situation
is similar to the muon-pair production
process: there is no interference between $s$-channel
$\ti{\nu}_{\tau}$ exchange, given by 
\begin{eqnarray}
G^s_{LL}=G^s_{RR}=\lambda_{131}\lambda'_{3kk}
\end{eqnarray}
and the SM $\gamma,Z$ processes, with the generalized charges
\begin{eqnarray}
Q^s_{ij}=-Q^q +g_i^eg_{-j}^q \frac{s}{s-m^2_Z+i\Gamma_Zm_Z}
\end{eqnarray}
while all other $Q_{ij}$ and $G_{ij}$ vanish. 
The unequal-flavor
down-type quark-pair production process, $e^+e^-\ra
d_j\bar{d}_k$, can be generated only by $s$-channel sneutrino with 
$G^s_{LL}=G^s_{RR}=\lambda_{131}\lambda'_{3jk}$.

\begin{figure}[htbp] 
\unitlength 0.8mm
\begin{picture}(80,150)
  \put(-10,-55){ \epsfxsize=13cm \epsfysize=14cm \epsfbox{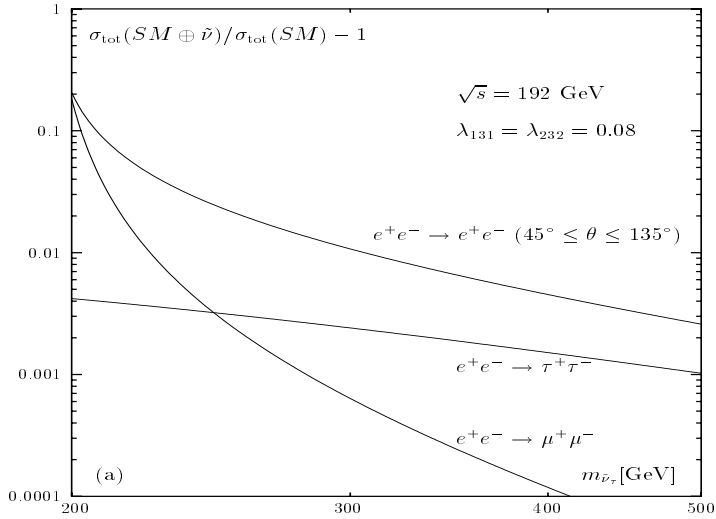}}
\end{picture}
\caption{
  Effect of sneutrino $\ti{\nu}_{\tau}$ exchange on the cross section
  for Bhabha scattering (restricting $45^{\circ} \leq \theta \leq
  135^{\circ}$), and $\mu^+\mu^-$ and $\tau^+\tau^-$ production at
  $\protect\sqrt{s}=192$ GeV.  }
\end{figure}
\begin{figure}[htbp] 
\unitlength 0.8mm
\begin{picture}(80,150)
  \put(-10,-55){ \epsfxsize=13cm \epsfysize=14cm\epsfbox{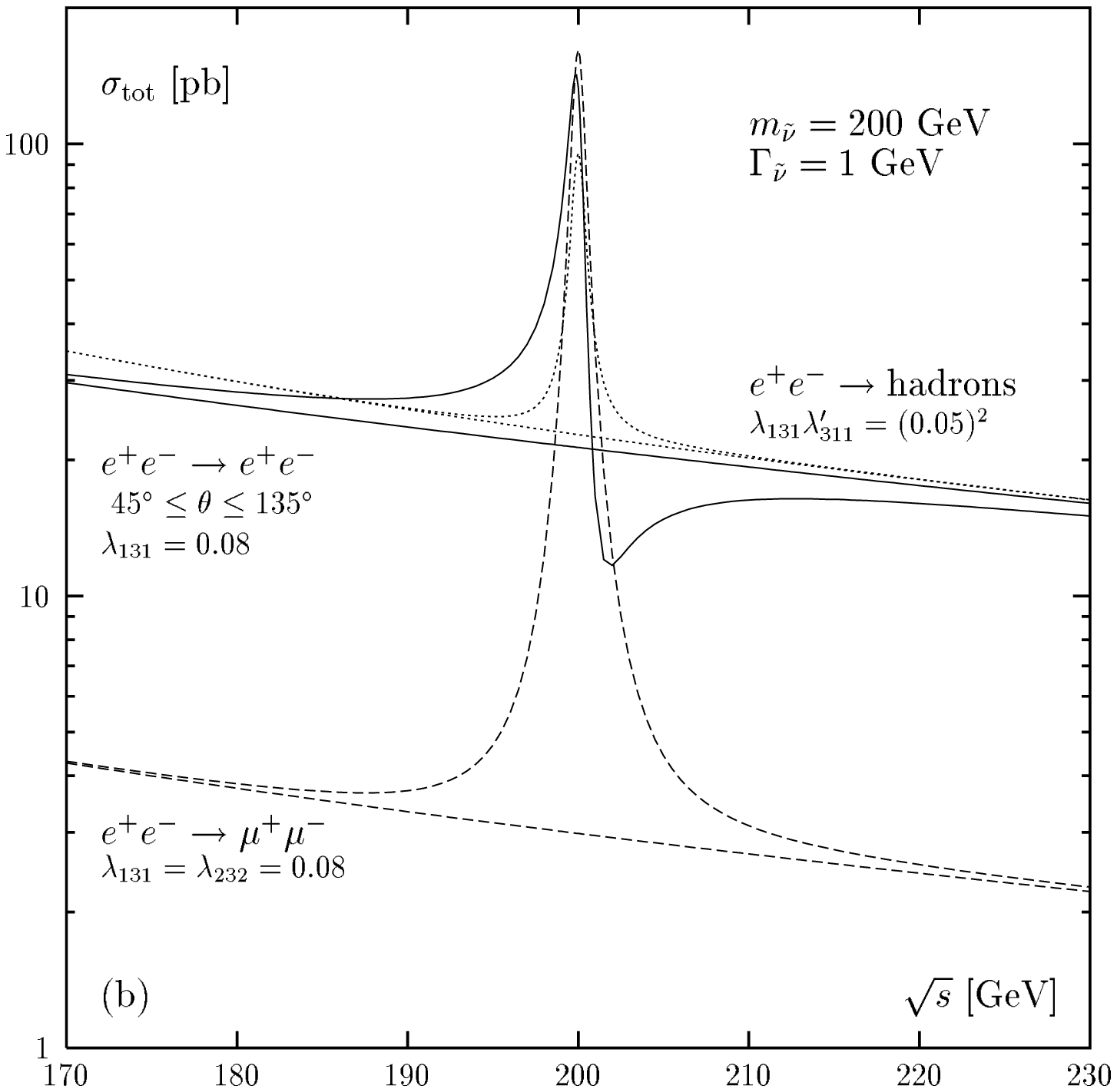}}
\end{picture}
\caption{ Cross section for Bhabha scattering
  (solid lines), $\mu^+\mu^-$ (dashed lines) and hadron production
  (dotted lines) in the SM, and including $\ti{\nu}_{\tau}$ sneutrino
  resonance formation as a function of the $e^+e^-$ energy. }
\end{figure}

Numerically the effect of $t$- or $u$-channel exchange of
sleptons is very small (typically below 1\%) for the slepton masses and
couplings consistent with low-energy data. Only in processes
with $s$-channel exchanges, and not too far from the resonance,
the effect of sneutrino can be quite spectacular. This is
illustrated in Fig.~3, where 
the impact of sneutrino $\ti{\nu}_{\tau}$ exchange on processes (a--c)
at LEP2 energy is shown. In the Figure the change of the total
cross section, $\sigma_{\rm tot}(SM\oplus \tilde{\nu})\ /\sigma_{\rm
  tot}(SM)-1$, is plotted as a function of sneutrino mass assuming
$\lambda_{i3i}=0.08$ and $\lambda_{131}\lambda'_{3jk}=(0.05)^2$ (for
Bhabha scattering the scattering angle is restricted to $45^{\circ}
\leq \theta \leq 135^{\circ}$). In processes (a) and (b), where the
$s$-channel sneutrino exchange can contribute, the effect can be very
large for sneutrino mass close to the LEP2 center-of-mass energy. Note
the difference due to different interference pattern between Bhabha
scattering and tau-pair production on the one hand, and muon-pair
production processes on the other: Bhabha and tau-production processes
are more sensitive to heavy sneutrinos.  If sneutrino is within the
reach of LEP2, a spectacular resonance can be observed in Bhabha
scattering, muon-pair, and/or quark-pair production processes, Fig.4;
again different interference patterns are seen. In the calculations
the total decay width $\Gamma_{\ti{\nu}_{\tau}}=1$ GeV has been
assumed.  Although the partial decay width $\Gamma(\ti{\nu}_{\tau}\ra
e^+e^-)=\lambda_{131} m_{\ti{\nu}_{\tau}}/16\pi$ is very small,
sneutrinos can also decay via $R$-parity conserving couplings
to $\nu\chi^0$ and $\ell^\pm\chi^\mp$ pairs with subsequent $\chi^0$
and $\chi^\pm$ decays. The partial decay widths into these channels
depend on the choice of supersymmetry breaking parameters,
however we find that in large
regions of the parameter space the total decay width can  be
as large as 1 GeV. Therefore it is significantly larger than the energy
spread $\delta E \sim 200$ MeV at LEP2 and in such a case the interference
with the SM processes must be taken into account. The peak cross
section for Bhabha scattering is given by the unitarity limit
$\sigma_{peak}=8\pi B_e^2/m^2_{\ti{\nu}_{\tau}}$ with sneutrino and
antisneutrino production added up, where $B_e$ is the branching ration
for the sneutrino decay to $e^+e^-$. An interesting situation may
occur if sneutrinos mix and mass eigenstates are split by a few GeV
\cite{snumix}.  Then one may expect to observe in the energy
dependence in Fig.4 
for the processes (a), (b) and/or (d) two separated peaks with reduced
maximum cross sections.

The angular distribution of leptons and quark jets is nearly isotropic
on the sneutrino resonance. As a result, the strong forward-backward
asymmetry in the Standard Model continuum is reduced to $\sim 0.03$ on
top of the sneutrino resonance. The deviations of the Bhabha cross
section from the SM expectations would allow to determine directly the
$\lambda_{131}$ coupling, or to derive an upper limit.  Similarly from
the other processes one could derive limits for $\lambda_{232}$ and
$\lambda'_{3jk}$. For example, if the total hadronic cross section at
192 GeV can be measured to an accuracy of 1\%, the Yukawa couplings
for a 200 GeV sneutrino can be bounded to
$\lambda_{131}\lambda'_{311}\lsim (0.045)^2$ \cite{tev}.  Recently preliminary
  results for some of the couplings from LEP 172 GeV data have been
  published \cite{lepres}.

\section{Sleptons at Tevatron}
Sleptons can also manifest themselves in $p\bar{p}$ collisions. 
At the Tevatron the case $\lambda'_{311}$ is the most
interesting since it allows for $\ti{\nu}_{\tau}$ and $\ti{\tau}$ 
resonance formation in valence quark collisions. 
Even though the sneutrinos and charged sleptons are expected to
have small widths of the order $\sim 1$ GeV, 
their decays to quark jets can be very difficult to observe in
the hadronic environment. 
Therefore we will consider leptonic decays of sleptons via
$\lambda_{i3i}$ couplings. To be specific we will consider
$\lambda_{131}$ and discuss $e^+e^-$ and $e^+\nu_e$ production
in $p\bar{p}$ collisions; the same results hold for $\mu^+\mu^-$
and $\mu^+\nu_{\mu}$ production if $\lambda_{232}$ is
assumed.

The differential cross
sections for $p\bar{p}\ra e^+e^-$ and $e^+\nu_e$ processes 
are obtained by combining the parton cross sections
with the luminosity spectra for quark-antiquark annihilation
\begin{eqnarray}
\frac{\mbox{d}^2\sigma}{\mbox{d}M_{\ell\ell}\mbox{d}y} 
[p\bar{p}\ra \ell_1\ell_2]=\sum_{ij} \frac{1}{1+\delta_{ij}}
\, \left(f_{i/p}(x_1)  f_{j/\bar{p}}(x_2) +(i 
\leftrightarrow j)\right)\,  \hat{\sigma}  
\label{lum}
\end{eqnarray}
$\hat{\sigma}$ is the cross section for the partonic subprocess
$ij\ra\ell_1\ell_2$, where $\ell_1\ell_2=e^+e^-$ or $e^+\nu$,
and 
$x_1=\sqrt{\tau}e^y$, $x_2=\sqrt{\tau}e^{-y}$.  $M_{\ell\ell}=(\tau
s)^{1/2} = (\hat{s})^{1/2}$ is the mass and $y$ the rapidity of the
lepton pair.  The probability to find a parton $i$ with momentum
fraction $x_i$ in the (anti)proton is denoted by
$f_{i/p(\bar{p})}(x_i)$.

The partonic differential cross sections in the
$q\bar{q}^{(')}$ center-of-mass 
frame are given by eq.~(\ref{dsigdcos}) with $A_c=1/3$, and $s$,
$t$ and $u$ replaced by $\hat{s}$, $\hat{t}$ and $\hat{u}$ which
refer to the $q\bar{q}^{(')}\ra \ell_1\ell_2$ subprocess. 
The $e^+e^-$ and $e^+\nu_e$ production processes are specified
as follows\\[1mm]
\noindent (a) {\it The process $q\bar{q}\ra e^+e^-$:} 
The SM $\gamma$ and $Z$ exchange mechanisms are given by  the 
generalized charges are as follows
\begin{eqnarray}
Q^s_{ij}=-Q^q+ g^q_L g^e_{-j}\frac{\hat{s}}{\hat{s}-m^2_Z
+i\Gamma_Z m_Z}
\end{eqnarray}
On the other hand, the $s$-channel sneutrino $\ti{\nu}_{\tau}$ 
exchange contributes
only to $d\bar{d}$ scattering with 
\begin{eqnarray}
G^s_{LL}=G^s_{RR}=\lambda_{131}\lambda'_{311}
\end{eqnarray}
which does not interfere with  the SM $s$-channel $\gamma,Z$
processes. 
All other $Q_{ij}$ and $G_{ij}$ vanish.\\[1mm]
\noindent (b) {\it The process $u\bar{d}\ra e^+\nu_e$:} 
This process proceeds via the $s$-channel $W$ boson and
$s$-channel $\ti{\tau}$ slepton exchanges. Only
\begin{eqnarray}
&&Q^s_{LR}=\frac{1}{2\sin^2\theta_W}\,\frac{\hat{s}}
{\hat{s}-m^2_W+i\Gamma_Wm_W}\\
&&G^s_{LL} =-\lambda_{131}\lambda'_{311}
\end{eqnarray}
are non-zero; all other $Q_{ij}$ and $G_{ij}$
vanish. 

\begin{figure}[htbp] 
\unitlength 0.8mm
\begin{picture}(80,150)
  \put(-10,-55){ \epsfxsize=13cm \epsfysize=14cm \epsfbox{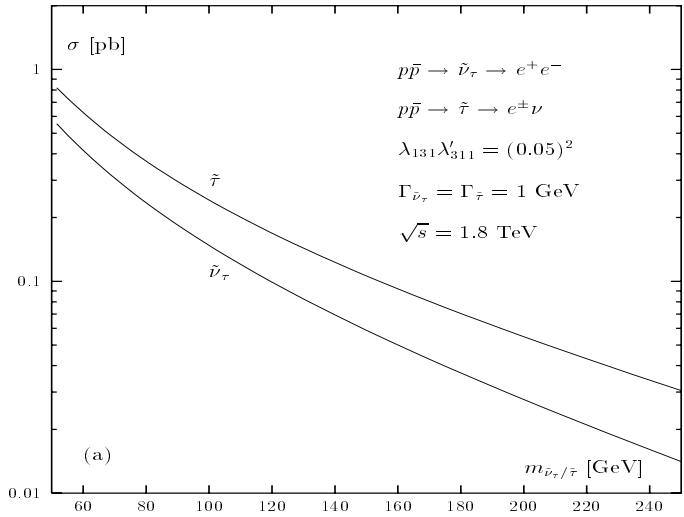}}
\end{picture}
\caption{ The cross section for sneutrino and antisneutrino 
  ($\ti{\nu}_{\tau}$) and stau ($\ti{\tau}$) production at the
  Tevatron, including the branching ratios to lepton-pair decays.  }
\end{figure}
\begin{figure}[htbp] 
\unitlength 0.8mm
\begin{picture}(80,150)
  \put(-10,-55){ \epsfxsize=13cm \epsfysize=14cm \epsfbox{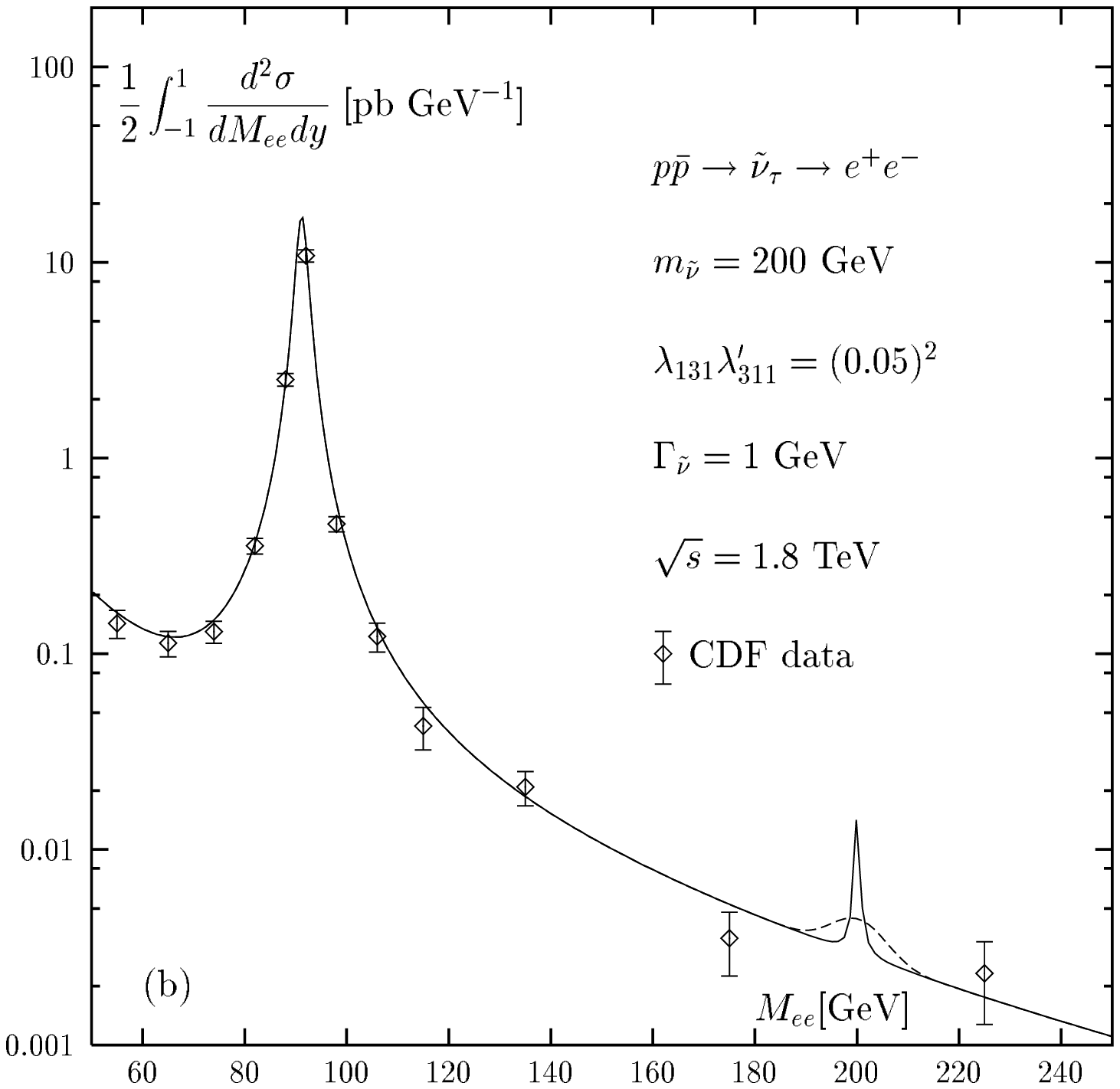}}
\end{picture}
\caption{The $e^+e^-$ invariant mass distribution including the $s$-channel
  sneutrino in the channel $d\bar{d}\ra e^+e^-$ is compared with the
  CDF data; solid line: ideal detector, dashed line: sneutrino
  resonance smeared by a Gaussian width 5 GeV. The CTEQ3L structure
  functions have been used. }
\end{figure}

The total cross sections for $\ti{\nu}_{\tau}$ and $\ti{\tau}$  production
in $e^+e^-$ and $e^+\nu_e$ channels, respectively, at Tevatron are
shown in Fig.5 as a function of the corresponding slepton mass. 
The total decay widths
of sleptons have been set to a typical value of 1 GeV, corresponding
to the branching ratios for leptonic decays of order 1\%. The
di-electron invariant mass distribution is compared to the CDF data in
Fig.6, where, following CDF procedure \cite{CDF}, the prediction for
$\frac{1}{2}\int^{1}_{-1}
{\mbox{d}^2\sigma}/{\mbox{d}M_{ee}\mbox{d}y}$ is shown. The
solid line corresponds to an ideal detector, while the dashed curve
demonstrates the distribution after the smearing of the peak by
experimental resolution characterized by a Gaussian width of 5 GeV. In
both plots the CTEQ3L parametrization \cite{cteq} is used together
with a multiplicative $K$ factor for higher order QCD corrections to
the SM Drell-Yan pair production.  The corresponding $K$ factor for
slepton production has not been calculated yet, leading to a
theoretical uncertainty in the $\lambda\lambda'$ couplings at a level
of about 10\%.
Assuming the sneutrino contribution to be smaller than the
experimental errors, we estimate that the bound
$\lambda_{131}\lambda'_{311}\lsim (0.08)^2 \ti{\Gamma}^{1/2}$ can be
established \cite{tev}, where $\ti{\Gamma}$ denotes the sneutrino width in
units of GeV.

\section{Summary}
The $R$-parity violating formulation of MSSM offers a distinct
phenomenology and therefore deserves detailed studies. Even if the
squarks are beyond the kinematical reach of HERA, sleptons might be
light enough to be seen at LEP2 and/or Tevatron. In this talk we
discussed the scenario with lepton number violation,  and we 
considered a number of processes in which sleptons might play an important
role. We concentrated only on those processes in which sleptons are
produced and decay via $R_p$ violating couplings. If the lepton-flavor
violating couplings are close to current low-energy limits, and the
slepton masses are in the range of 200 GeV, spectacular events can be
expected at both LEP2 and Tevatron. On the other hand, if no
deviations from the SM expectations are observed, stringent bounds on
individual couplings can be derived experimentally in a direct way.

\section*{Acknowledgments}
It is a pleasure to thank K. Ko\l odziej 
for the invitation and warm hospitality at the school.
I am grateful to R.~R\"uckl, H.~Spiesberger and P.~Zerwas
for their collaboration and comments on the manuscript. I would also
like to thank H.~Dreiner, 
W.~Krasny,  Y.~Sirois and F.~\.Zarnecki for
discussions and communications.


\vspace{-14pt}
\begin{thebibliography}{99}
\bibitem{data} H1 Collab., C. Adloff et al., \zpc74,97,191;\\
   Zeus Collab., J. Breitweg et al., \zpc74,97,207
\bibitem{spi} talks by H. Spiesberger, and by F. Cornet, 
   these proceedings and references therein
\bibitem{FF} G. Farrar and P. Fayet, \plb76,78,575
\bibitem{WSY} S.  Weinberg, \prd26,82,287; \\ 
    N. Sakai and T. Yanagida, \npb197,82,533  
\bibitem{D1} H. Dreiner, hep-ph/9707435
\bibitem{pdb} R.M. Barnett et al., PDG, \prd54,96,1
\bibitem{snu}J.~Kalinowski, R.~R\"uckl, H.~Spiesberger and
  P.M.~Zerwas, \plb406,97,314
\bibitem{tev} J.~Kalinowski, R.~R\"uckl, H.~Spiesberger and
  P.M.~Zerwas, DESY 97-062, hep-ph/9708272
\bibitem{lq} J.~Kalinowski, R.~R\"uckl, H.~Spiesberger and
  P.M.~Zerwas, \zpc74,97,595 
\bibitem{limits} For an update on experimental limits, see H. Dreiner
  in \cite{D1}, and G.~Bhattacharyya, Proc.\ 
   ``Beyond the Desert 97'', Ringberg Castle, June 1997
\bibitem{barger} V.\ Barger, G.F.\ Giudice and T.\ Han, \prd40,89,
  2987 
\bibitem{agashe} K.~Agashe and M.~Graessler, \prd54,95,4445
\bibitem{deb} G.~Bhattacharyya and D.~Choudhury, \mpl10,95,1699
\bibitem{dl} R.~M.~Godbole, P.~Roy and X.~Tata, \npb401,93, 67;\\
  H.~Dreiner and S.~Lola, Proceedings, ``$e^+e^-$ Collisions
  at TeV Energies: The Physics Potential", Annecy--Gran Sasso--Hamburg
  Workshop 1996, DESY 96-123D
\bibitem{lepres} M. Pohl, L3 Collab., presented at the LEPC Meeting,
  May 29, 1997;\\
  papers submitted to the XVIII Int.\ Symposium on Lepton Photon
Interactions:  Delphi Collab., LP-271; Aleph Collab., LP-278; 
Opal Collab., LP-313
\bibitem{snumix} M. Hirsch, H.V.  Klapdor-Kleingrothaus and S.G.
  Kovalenko, hep-ph/9701253; hep-ph/9701273;\\
  Y. Grossman and H.E. Haber, SLAC-PUB-7423, hep-ph/9702421
\bibitem{CDF} F. Abe et al., Fermilab-Pub-97/171-E, to appear in {\it
    Phys.\ Rev.\ Letters}
\bibitem{cteq} H.L. Lai et al., \prd51,95,4763

\end{thebibliography}
\end{document}